\documentclass[a4paper,10pt,notitlepage]{article}

\newcommand{\Dy}{$(\mathrm{Dy_{47},Nd_{53}})_{2}\mathrm{Fe}_{14}\mathrm{B}$ } 
\newcommand{\Nd}{$\mathrm{Nd}_{2}\mathrm{Fe}_{14}\mathrm{B}$ } 
\newcommand{\PrFeB}{$\mathrm{Pr}_{2}\mathrm{Fe}_{14}\mathrm{B}$ } 
\newcommand{\kT}{$25\,\mathrm{ k_{B}}T$}

\usepackage[utf8]{inputenc}
\usepackage{graphicx}

\usepackage{units}
\usepackage{amstext}
\usepackage{amssymb}

\usepackage{authblk}

\newcommand\blfootnote[1]{%
  \begingroup
  \renewcommand\thefootnote{}\footnote{#1}%
  \addtocounter{footnote}{-1}%
  \endgroup
}

\begin{document}

\title{Thermal activation in permanent magnets} 

%
%
%
%
%

\author[1]{S. Bance\thanks{s.g.bance@gmail.com, http://academic.bancey.com}}
\author[1]{J. Fischbacher}
\author[1]{A. Kovacs}
\author[1]{H. Oezelt}
\author[1]{F. Reichel}
\author[1,2]{T. Schrefl}
\affil[1]{Department of Technology, St. P\"{o}lten University of Applied Sciences, A-3100 St P\"{o}lten, Austria}
\affil[2]{Center for Integrated Sensor Systems, Danube University Krems, 2700 Wiener Neustadt, Austria}

\date{21 April 2015}


\maketitle 

\blfootnote{
This article is published in JOM ``The Journal of The Minerals, Metals \& Materials Society'' as 
S. Bance et al., ``Thermal Activation in Permanent Magnets`` JOM, Volume 67, Issue 6, pp 1350-1356 (April 2015) doi:10.1007/s11837-015-1415-7; http://link.springer.com/article/10.1007/s11837-015-1415-7}

\begin{abstract}
The coercive field of permanent magnets decays with temperature. At non-zero temperature the system can overcome a finite energy barrier through thermal fluctuations. 
Using finite element micromagnetic simulations, we quantify this effect, 
which reduces coercivity in addition to the decrease of the coercive field associated with the temperature dependence of the anisotropy field, 
and validate the method through comparison with existing experimental data. 

\end{abstract}

\clearpage

\section{Introduction}
The importance of permanent magnets to many modern technologies has led to increased interest in developing magnets that contain lower amounts of supply-critical materials \cite{Gutfleisch2011}. 
Progress in processing, characterization and simulation of rare earth permanent magnets has helped continually improve their performance. 
Reversal depends on the microstructure, with the grain boundary phase and surface defects being of particular importance. 
Numerical micromagnetics is successfully being used to understand the reversal mechanisms that determine important extrinsic properties such as the coercive field $H_{\mathrm{c}}$ \cite{bance2014influence}. 

Understanding the temperature dependence of coercivity is of importance in the design of permanent magnets for applications at high temperatures, 
e.g. in the motors of electric and hybrid vehicles where the operating temperature is typically around $T=450\mathrm{\,K}$.  
In leading order, the temperature dependence of the intrinsic magnetic properties causes the reduction of the coercive field with temperature, $T$ \cite{Skomski2013}. 
This is expressed by the well known relation \cite{Kronmueller1988}
\begin{equation}
H_{\mathrm c}(T) = \alpha \frac{2K(T)}{\mu_0 M_{\mathrm s}(T)} - N_{\mathrm{eff}} M_{\mathrm s}(T)
\label{eq:kroni}
\end{equation}
which relates the coercive field, $H_{\mathrm{c}}$, to the  the anisotropy constant, $K(T)$, and the magnetization, 
$M_{\mathrm s}(T)$. In (\ref{eq:kroni}) $\mu_0$ is the permeability of vacuum. While equation (\ref{eq:kroni}) is widely used to classify permanent magnets based on the microstructural parameters $\alpha$ and $N_{\mathrm{eff}}$, it also expresses the main contribution to the temperature dependence of $H_{\mathrm c}(T)$. In particular equation  (\ref{eq:kroni}) relates the coercive field to the nucleation field, ${2K(T)}/({\mu_0 M_{\mathrm s}(T)})$, of a small magnetic sphere \cite{Brown1959}.
The role of thermal fluctuations becomes evident through viscosity experiments. Under the action of a constant applied field the magnetization decays with time \cite{Wohlfarth1984}. The change of magnetization within the time $t$ is given by
\begin{equation}
\label{eq:viscosity}
\Delta M(t) = - S \ln(t).  
\end{equation}
The viscosity, $S$, is attributed to irreversible changes of the magnetization across the
energy barrier. The logarithmic dependence results from the distribution of energy barriers in the magnet. 
Under the assumption that a coercivity is related to the expansion of an already reversed nucleus the
coercive field of a permanent magnet can be written as \cite{Givord1988,barthem2002analysis}
\begin{equation}
\label{eq:givord}
H_{\mathrm c}  = \alpha' \frac{\gamma}{\mu_0 M_{\mathrm s}v^{1/3}} - N_{\mathrm{eff}} M_{\mathrm s}  
 - \frac{25 k_{\mathrm B}T}{\mu_0 M_{\mathrm s}v}. 
\end{equation}
Here $k_{\mathrm B}T$ is the Boltzmann constant and $\alpha'$ replaces $\alpha$. Similarly, to equation (\ref{eq:kroni}) the intrinsic parameters and the derived quantities depend on temperature. In order to improve the readability, we have dropped the $(T)$ behind the symbols. $\gamma = \gamma(T)$ is the energy per unit area of a Bloch wall, $\gamma = 4\sqrt{AK}$, with the exchange constant $A = A(T)$.  The activation volume, $v = v(T)$, may be associated with the volume of the initial nucleus. The last term in equation (\ref{eq:givord}) is proportional to the fluctuation field. The fluctuation field drives the systems over an energy barrier of $\Delta E = 25 k_{\mathrm B}T$ within the characteristic measurement time. This energy can be overcome within the characteristic measurement time of the coercive field. Applying the Arrhenius-N\'{e}el law the relaxation time over an energy barrier, $\Delta E$ is
\begin{equation}
\tau =\frac{1}{f_{0}}\mathrm{exp}\Bigg(\frac{\Delta E}{k_{\mathrm{B}}T}\Bigg)
\label{equation:arrenhius}
\end{equation}
where $f_{0}$ is the attempt frequency, which limits the probability for reversal. 
The first term of equation (\ref{eq:givord}) can be rewritten as $\alpha''{2K(T)}/{(\mu_0 M_{\mathrm s})}$ when the
activation volume is assumed to be proportional to the Bloch wall width $\delta_{\mathrm B} = \sqrt{A/K}$.   
The parameters $\alpha$, $\alpha'$, and $N_{\mathrm{eff}}$ can be derived by fitting the measured temperature dependent coercive field to equations (\ref{eq:kroni}) and (\ref{eq:givord}) \cite{kou1994coercivity}. 

In particulate and thin film recording the particle or grain size is small. The total magnetic volume, $V$, is low. In zero field the energy barrier for magnetization reversal is given by the smaller of the two values $\Delta E_{0} = KV$ or $\Delta E_{0} = 4F\sqrt{AK}$, where $F$ is the minimum cross section of a columnar grain. At an opposing field $H$ the energy barriers \cite{sharrock1990time} decays with field according to
\begin{equation}
\Delta E = \Delta E_{0}\big(1-H/H_{\mathrm{0}}\big)^{n}.
\label{equation:sharrock}
\end{equation}
Here $\Delta E_{0}$ is the energy barrier in zero field and $H_0$ is the field where the barrier is zero. 
The exponent $n$ has been discussed in detail in the literature, covering the dependence of $n$ on factors including external field strength and field angle \cite{harrell_orientation_2001, suess_reliability_2007}. 
A value close to $n=2$ is usually used in situations corresponding to coherent reversal, while $n=1.5$ corresponds to nucleation and expansion. 
When using micromagnetics, as in this article, the reversal mode is known directly from the calculations, so it is not necessary to know the value of $n$ to determine the type of reversal mechanism. 
Equations (\ref{equation:arrenhius}) and (\ref{equation:sharrock}) lead to a time and temperature dependent
coercive field \cite{sharrock1990time}
\begin{equation}
H_{\mathrm c}(t,T) = H_0\left(1-\left(\frac{k_{\mathrm B}T}{\Delta E_{0}}\ln(f_0t) \right)^{1/n} \right) 
\label{equation:sharrock2}
\end{equation}
Equation (\ref{equation:sharrock2}) gives the field that causes switching of half of the particles or grains within time $t$.

Because of the much larger grain size in permanent magnets and higher magneto-crystalline anisotropy in modern permanent magnets as compared to magnetic recording materials, it was widely believed that thermal fluctuations play only a minor role during magnetization reversal in permanent magnets. In permanent magnets the magnetization reversal is initiated within a small volume: either the reversed nucleus or the volume associated with a domain wall depinning process. Similarly to magnetization reversal in small particles thermal activation helps to initiate reversal within this characteristic volume.  Advances in computational methods and computing power have made it possible to compute the effects of temperature in permanent magnets taking into account both the temperature dependent intrinsic properties and the thermal fluctuations over finite energy barriers. 
Using methods from chemical physics \cite{henkelman2000climbing}, we compute the energy barrier as a function of the field and thus can estimate the influence of thermal fluctuations on the coercive field. Details of the computations will be presented in section \ref{sec:method} of this paper. We will demonstrate the influence of thermal fluctuations on coercivity for Pr$_2$Fe$_{14}$B magnets in section \ref{sec:cuberesults}, comparing the angle dependence of coercivity at 4.2 K, 175 K and 300 K.

Traditional Nd\(_{2}\)Fe\(_{14}\)B permanent magnets are doped with dysprosium to improve their performance. 
The higher uniaxial anisotropy $K$ and lower magnetization $M_{\mathrm{s}}$ of the dysprosium increases the anisotropy field $H_{\mathrm A}=2K/(\mu_0M_{\mathrm{s}})$,  which results in a higher $H_{c}$ but, of course, a lower overall magnetization. 
Importantly, the market price of dysprosium and other heavy rare earth elements peaked drastically in 2010, 
prompting a frantic search for new permanent magnets using cheaper materials. In order to produce magnets with high energy product, using fewer rare earth elements, a number of routes are currently being followed. 
In addition to the grain boundary phase which separates the grains magnetically, modern magnets use the concept of magnetic  surface hardening \cite{Ghandehari1987,Nakamura2005} for improved coercivity. The local anisotropy field near the surfaces of each grain is increased by partially substituting Nd with Dy in Nd\(_{2}\)Fe\(_{14}\)B based magnets. In section \ref{sec:dodecresults} we compute the coercivity 
of Nd$_2$Fe$_{14}$B grains with a thin (Dy,Nd)$_2$Fe$_{14}$B shell.

\section{Method}
\label{sec:method}

In this work we follow a computational micromagnetics approach to treat the temperature dependence of the coercive field. The classical nucleation field theory starts \cite{Brown1959} from the uniform magnetic states and determines the critical field when this state becomes unstable. Under the action of an opposing field the system is in a metastable minimum. An energy barrier separates this local minimum (magnetization and field antiparallel) of the global minimum (magnetization and field parallel). With increasing opposing field the energy barrier decreases. At the nucleation field the local minimum vanishes; the system is at a saddle point and may reverse towards a global minimum. 
In non-ellipsoidal particles the demagnetizing field is non-unform. In turn the remanent magnetic state and magnetic states under an opposing external field are inhomogeneous. However, a similar stability criterion as for the ellipsoid may be applied to define the nucleation field \cite{schabes1988magnetization,schabes1991micromagnetic,schmidts1994algorithm}. Standard numerical micromagnetic methods implicitly apply this criterion for the calculation of the switching field. The temperature dependence of the coercive field can be computed when the temperature dependent intrinsic magnetic properties, $M_{\mathrm s}(T)$, $K(T)$, and $A(T)$ are used as input for the computations. Similarly to an experiment these computed temperature dependent values for the coercive field can be fitted to equation (\ref{eq:kroni}) \cite{sepehri2014micromagnetic}.  

In addition to the effect of the temperature dependent intrinsic magnetic properties on the coercivity, the influence of the thermal fluctuations on the temperature dependence of the coercive field can be addressed using numerical micromagnetics. By means of methods widely applied in chemical physics for the computation of reaction rates \cite{henkelman2000climbing}, it is possible to compute the height of the energy barrier separating the local minimum associated with the magnetic state before the reversal from the minimum that corresponds to the reversed state \cite{dittrich2002path}. Reversal occurs when the opposing field reduces the energy barrier to a height that can be overcome by thermal energy \cite{dittrich2005thermally}. This field is the coercive field and is a function of temperature. This method has been applied to compute the finite temperature switching field of permalloy elements and the thickness dependence of the coercive field in granular recording media \cite{suess2011calculation}. 

\begin{figure*}[t]
\includegraphics[width=0.9\textwidth]{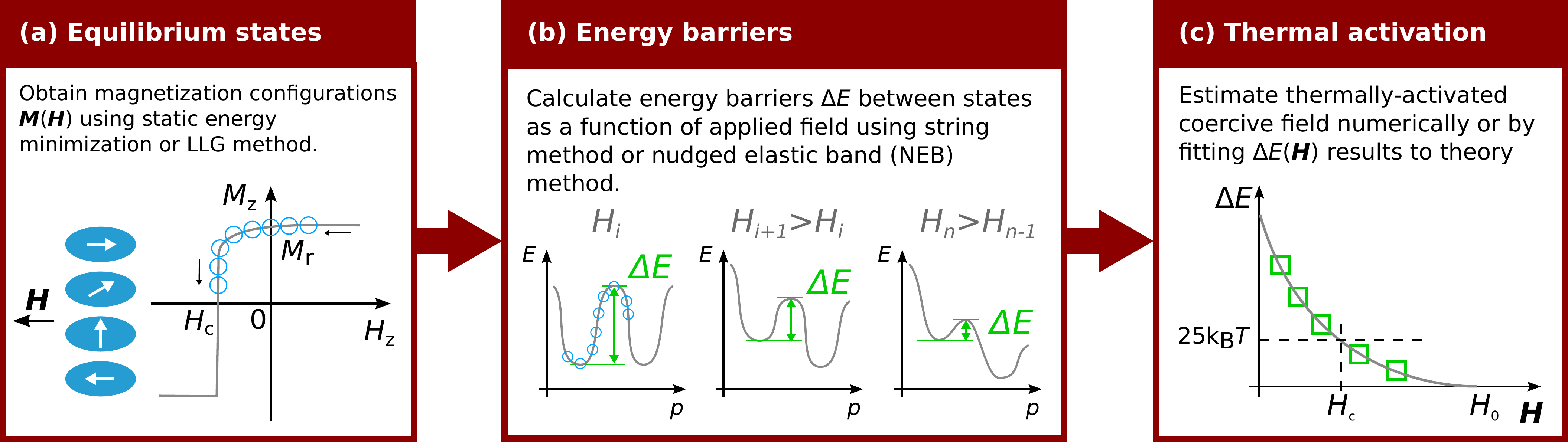}
\caption{Schematic representation of the method for calculating thermally activated coercivity using numerical micromagnetics: (a) equilibrium states are calculated during a hysteresis simulation, using LLG micromagnetics or an energy minimization method; (b) energy barriers along a path in configuration space are calculated using the elastic band method or string method. $p$ is the distance along this path; (c) the thermally activated coercivity is estimated as the field required to reduce the energy barrier height to \kT. }
\label{fig:fig1}
\end{figure*}

The simulation process is illustrated in Fig. \ref{fig:fig1}. We first compute the demagnetization curve of the magnet using a standard micromagnetic solver (Fig. \ref{fig:fig1}a). We can either compute minimum energy states \cite{exl2014} or solve the Landau-Lifshitz Gilbert (LLG) equation \cite{suess2002time} for different applied fields. The coercive field obtained for the simulation of the demagnetization curve corresponds to $H_0$ in equation (\ref{equation:sharrock2}). It is the field where the energy barrier that separating the states before and after irreversible switching is zero. Then we want to compute the energy barrier between a state with $|H_i| < |H_0|$ which we denote with $\mathbf M_{\mathrm {initial}}$ and the reversed state which is called $\mathbf M_{\mathrm {final}}$. The transient magnetic states from the computation of the demagnetization curve serve as initial path for the computation of the minimum energy path connecting $\mathbf M_{\mathrm {initial}}$ and $\mathbf M_{\mathrm {final}}$ (Fig. \ref{fig:fig1}b). A path is optimal, if for any point along the path the gradient of the energy is parallel to the path. In other words: the component of the energy gradient normal to the path is zero. This path is called the minimum energy path, which means that the energy is stationary for any degree of freedom perpendicular to the path. Let's denote the magnetic state of the point with the maximum energy in the path by $\mathbf M^*$. This is the saddle point. The difference between the energy of the saddle point and the initial state is the energy barrier:
\begin{equation}
\label{eq:barrier}
\Delta E(H_i) = E\left(\mathbf M^*\right) - E\left(\mathbf M_{\mathrm {initial}}(H_i)\right).
\end{equation}
We apply the climbing image nudged elastic band method \cite{henkelman2000climbing,dittrich2002path} or the modified string method \cite{E2007simplified} to compute the minimum energy path. Both methods take a series of magnetization configurations as input and minimize the energy path, formed from a series of connected nodes, in the multi-dimensional configuration space according to the local gradient at each node. They differ only in their algorithms. The nudged elastic band method employs a spring force between adjacent nodes in order that they do not become too separated. The string method renormalizes the distance between adjacent nodes after each iteration. We repeat the computation of the energy barrier (\ref{eq:barrier}) for different applied fields and fit the results to equation (\ref{equation:sharrock}). The critical field value at which the energy
barrier becomes 25~$k_{\mathrm B}T$ is the temperature dependent
coercive field, $H_{\mathrm c}(T)$, see Fig. \ref{fig:fig1}c and Fig. 4 in the paper by Sharrock \cite{sharrock1990time}.
The value of  25~$k_{\mathrm B}T$ is the energy barrier that follows from (\ref{equation:arrenhius}) for a typical measurement time of 1 second \cite{givord1987magnetic}. Hereby an attempt frequency of $f_0 = 10^{11} \mathrm{Hz}$ was assumed. The attempt frequency may depend on the nature of the domain nucleation or depinning process. Thus for a more accurate numerical treatment of $H_{\mathrm c}(T)$ a method for the computation of the attempt frequency such as forward flux sampling \cite{vogler2013simulating} may be applied.


\begin{table}[t]
\caption{Material properties of the phases used in the simulations. } 
\label{table:materials} 
\centering 
\begin{tabular}{c c c c c}
      \hline\hline 
      Name & $T$(K) & $K$(MJ/m\textsuperscript{3}) & $\mu_0M_{s}$(T) & $A$(pJ/m)\\  
      \hline 
      \PrFeB & 4.2 & 23.5\cite{Hirosawa1986} & 1.85\cite{Sagawa1985} & 11.3 \\ 
      \PrFeB & 175& 12.39\cite{Hirosawa1986} & 1.78\cite{Sagawa1985} & 10.6 \\ 
      \PrFeB & 300 & 5.40\cite{Hirosawa1986} & 1.56\cite{Sagawa1985} & 8.12 \\ 
      \Dy & 300 & 5.17\cite{Sagawa1987} & 1.151\cite{Sagawa1987} & 8.7\cite{Hawton1943}\\ 
      \Dy & 450 & 2.70\cite{Sagawa1987} & 0.990\cite{Sagawa1987} & 6.44\cite{Hawton1943}\\ 
      \Nd & 300 & 4.30\cite{Hock1988} & 1.613\cite{Hock1988} & 7.7\cite{Durst1986}\\      
      \Nd & 450 & 2.09\cite{Hock1988} & 1.285\cite{Hock1988} & 4.89\cite{Durst1986}\\      
      \hline 
\end{tabular}
\end{table}

We apply a finite element method for the computation of the demagnetization curve and the energy barriers. 
Rave and co-workers\cite{rave1998corners} suggest that the mesh size should be smaller than the theoretical exchange length, which is defined analytically as $L=\sqrt{A/\mu_0M_{\mathrm s}^2}$ for a ferromagnet. 
To satisfy this requirement while restraining the finite element mesh to a reasonable number of elements we use an adaptive mesh, 
where the fine mesh is constrained to the regions of domain wall nucleation. The intrinsic material constants used for the simulations are given in Table \ref{table:materials}.
The solver uses a hybrid finite element / boundary element method to calculate the external demagnetizing field and, unless otherwise stated, neighbouring phases in the models are fully exchange coupled according to the local material parameters. 

\section{Results \& Discussion}
\label{sec:results}

\subsection{Surface defects in PrFeB grains} 
\label{sec:cuberesults}
A single \PrFeB grain is modelled as a cube with 100 nm edge length and a soft surface defect with 0.8 nm thickness and uniaxial anisotropy constant $K=0$. 
Confining the defect to one corner allows a reduction in model size and thus simulation cost, since the nucleation region must be finely discretized, without changing the resulting critical fields. 
The initial magnetization is in the $+z$ direction, parallel to the c-axis. 
An opposing field is applied with an angle $\theta _{H}$ from the $-z$ direction in the $z-x$ plane.

\begin{figure}[t]
\includegraphics[width=1.0\columnwidth]{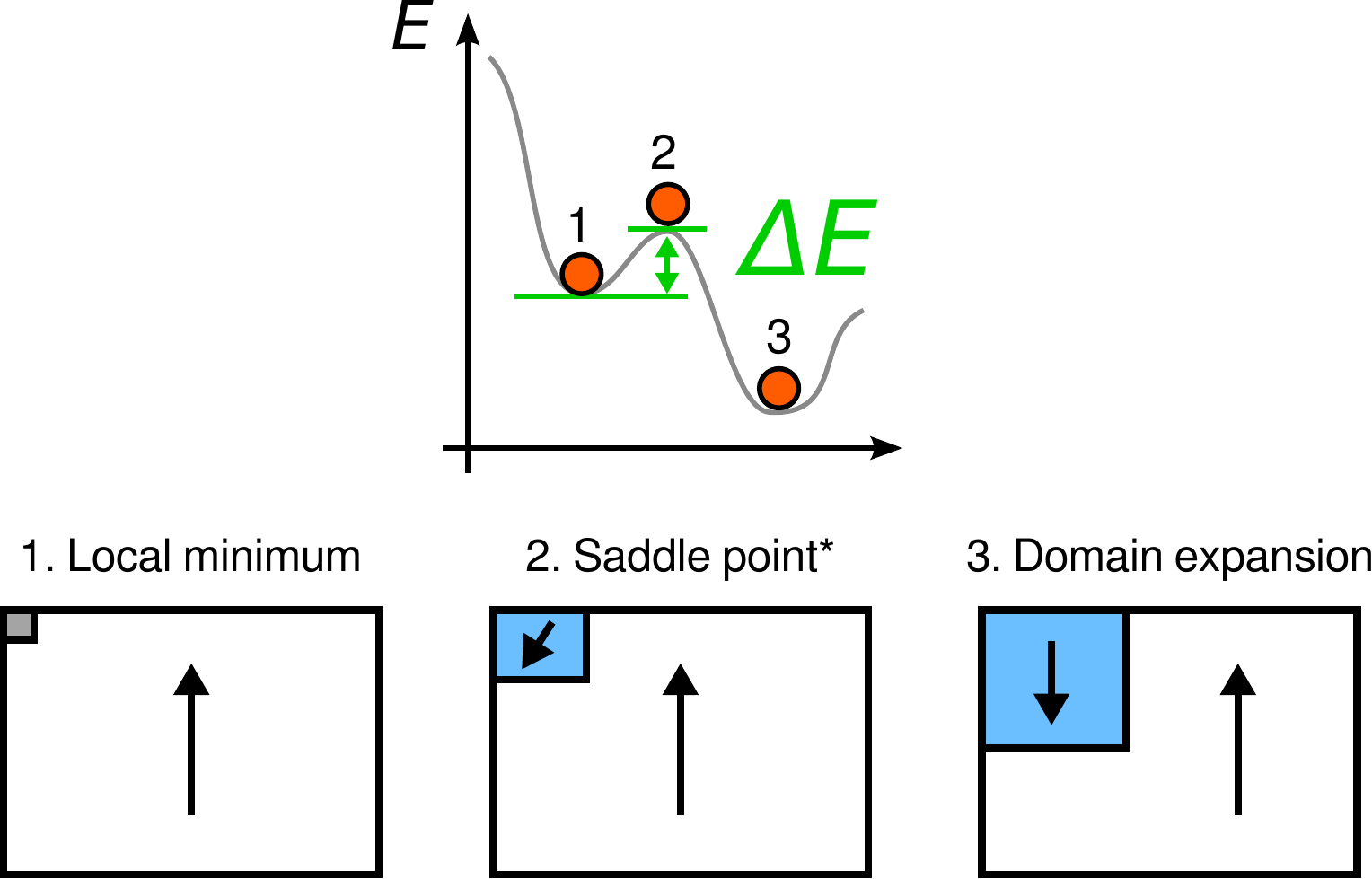}
\caption{Schematic of magnetization reversal. Sketches of the magnetic state in the metastable minimum before reversal (1), at the saddle point (2), and after passing the saddle point (3).}
\label{fig:reversal}
\end{figure}

Fig. \ref{fig:reversal} shows the thermally activated magnetization reversal process schematically.
Reversal begins with the rotation of the magnetization in the soft defect. 
As the field increases, the magnetization within the nucleus rotates towards the applied field direction. The Zeeman energy decreases and makes the presence of a domain wall energetically favourable. A domain wall like state forms between the nucleus and rest of the magnet. Increasing the field further reduces the energy barrier to zero, at which point the nucleus expands and the domain wall rapidly passes through the remaining grain volume. 

\begin{figure}[t]
\includegraphics[width=0.8\columnwidth]{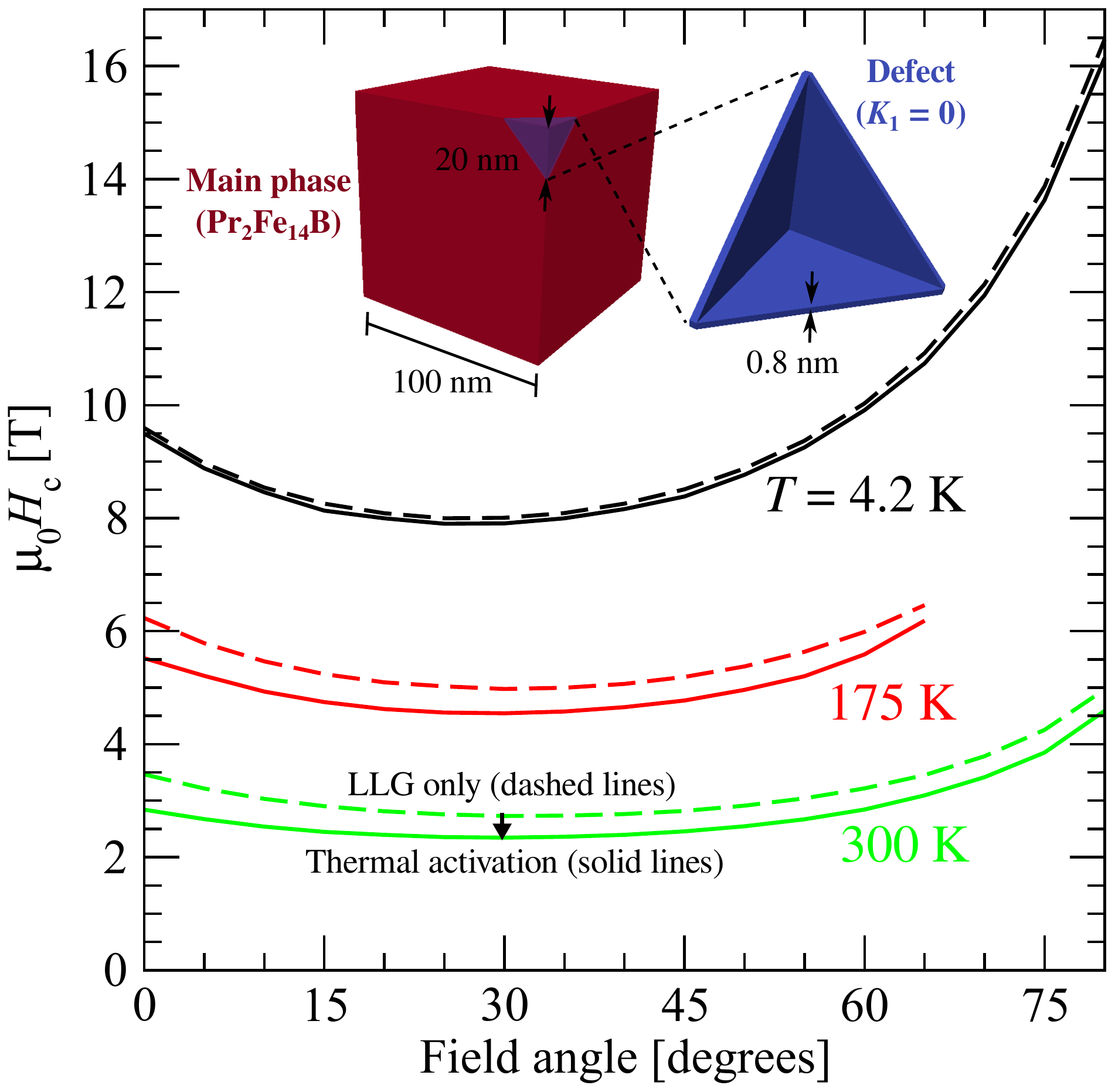}
\caption{Coercive field as a function of applied field angle for a cubic \PrFeB grain calculated without thermal activation (dashed lines) and including thermal activation (solid lines). 
The grain includes an anisotropy-reduced surface defect layer of 0.8 nm thickness and has an edge length of 100 nm. The inset images show the grain model used with the surface defect of 0.8 nm thickness and reduced uniaxial anisotropy. }
\label{fig:pr}
\end{figure}

The energy barriers corresponding to a range of applied field strengths are calculated by minimizing this path using the nudged elastic band method. Fitting the field-dependent barrier height $\Delta E(H_i)$ to equation (\ref{equation:sharrock}) the temperature dependent coercivity is estimated. Fig. \ref{fig:pr} shows the angle dependent coercive field computed for different temperatures. The dashed lines are computed with an LLG solver, which correspond to the temperature dependent coercivity, $H_0(T)$, taking into account the temperature dependent intrinsic parameters but neglecting possible fluctuations over finite barriers. The solid lines give the temperature dependent coercivity, $H_{\mathrm c}(T)$, including both the temperature dependent intrinsic material properties and thermal hopping over finite energy barriers. With increasing temperature the difference between $H_0(T)$ and $H_{\mathrm c}(T)$ increases. For a field angle of zero the relative change of the coercivity by thermal fluctuations
\begin{equation}
\Delta H_{\mathrm {fl}}(T) = \frac{H_0(T)-H_{\mathrm c}(T)}{H_0(T)}
\end{equation}
is 0.01, 0.11, and 0.18 for a temperature of 4.5~K, 175~K, and 300~K, respectively. 
It is interesting to note that with increasing temperature the minimum in the coercive field as function of temperature becomes less pronounced.  

\subsection{Grain boundary diffused magnetic grains} 
\label{sec:dodecresults}
\begin{figure}[t]
\includegraphics[width=0.95\columnwidth]{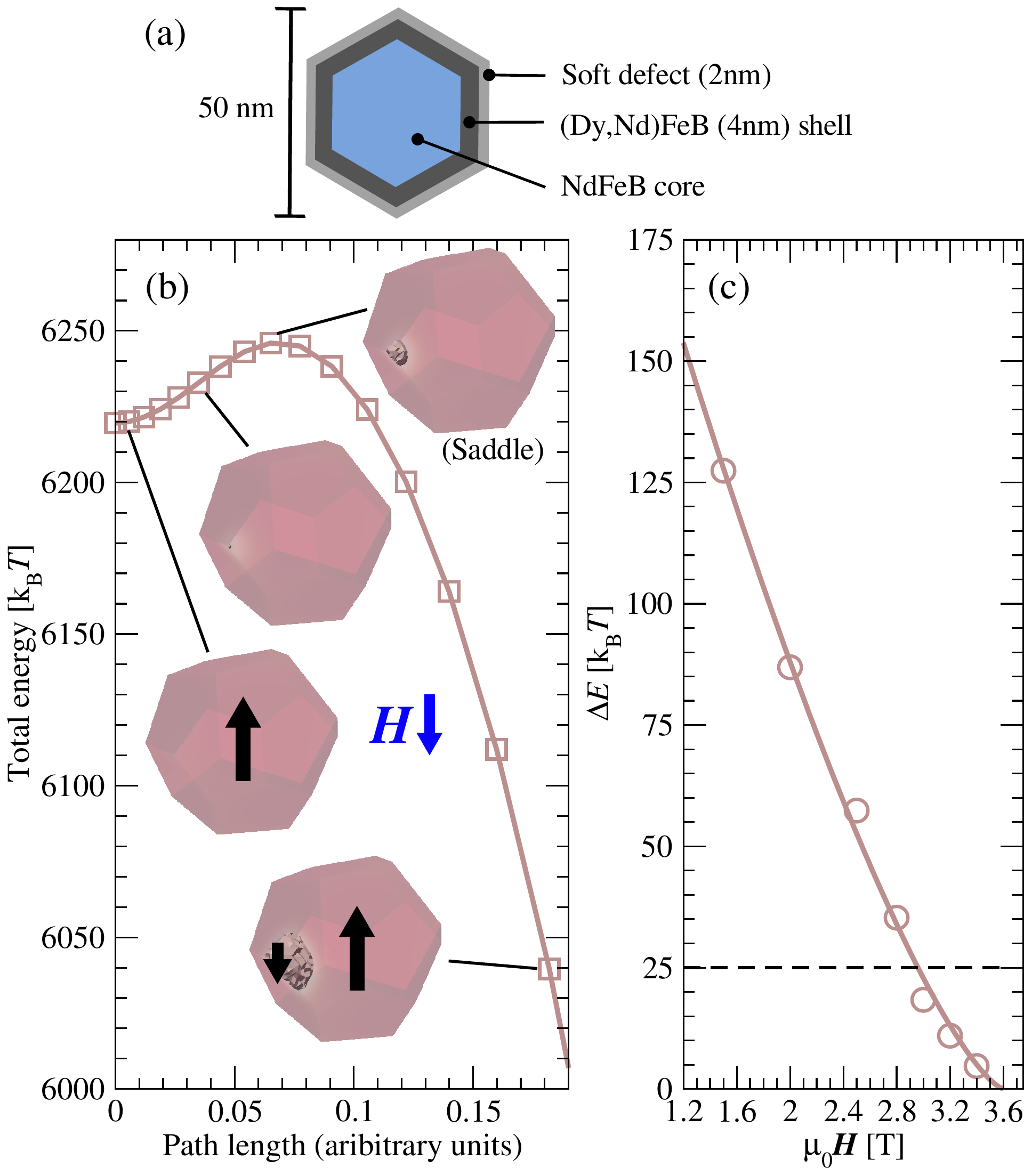}
\caption{Simulation results for calculation of thermally-activated coercive field in a single grain of a Dy grain boundary diffused core-shell permanent magnet at 450~K. 
(a) Schematic representation of the model (not to scale). 
(b) The minimum energy path during reversal under an applied field of 2.8 T, calculated using the string method. 
The saddle point of highest energy corresponds with the so-called activation volume. The energy barrier $\Delta E$ is the difference in the total magnetic Gibbs free energy between the initial configuration and the saddle configuration. 
The inset images show the nucleation of the reversal domain, which occurs at the outer surface, within the soft defect layer. Black arrows indicate the local magnetization direction, which initially opposes the applied field $H$. 
(c) $\Delta E$ is calculated for a range of applied field strengths, allowing an estimation of the field required to reduce the energy barrier to a height of \kT.} 
\label{fig:dy}
\end{figure}

We simulate a Dy grain boundary diffused magnetic grain with a \Nd core, a hard 4 nm \Dy shell and a 2 nm soft surface defect.
The soft defect has the properties of \Dy except the uniaxial magnetocrystalline anisotropy constant is reduced to $K=0$. 
The outer diameter of the dodecahedral grain is constant at 50 nm. 
Intrinsic material properties are given in Table \ref{table:materials}. The thermally activated reversal process at $T=450$~K is illustrated in Figure \ref{fig:dy}.
Reversal begins by rotation of the magnetic moments within the soft defect before nucleation of a reversal domain at the corner. 

The insets in Figure \ref{fig:dy}b show the grain and the formation of the reversed nucleus in the soft outer defect using a three-dimensional contour plot. 
The solid line plot gives the total energy of the system along the minimum energy path calculated using the string method with an applied field of 2.8 T. 
At the saddle point, the point with maximum energy along the minimum energy path, a small nucleus is formed at the corner of the dodecahedron. 
Figure \ref{fig:dy}c gives the energy barrier as function of the applied field, $\Delta E(H)$ along with the corresponding fit to Equation \ref{equation:sharrock}, where $n=1.37$. 
The critical field where the barrier crosses the line $\Delta E(H) = 25 k_{\mathrm B}T$ is the coercive field computed taking into account thermal fluctuations. 
Without thermal fluctuations the barrier would have to vanish ($\Delta E(H) = 0$) for magnetization reversal to occur. 
At a temperature of $T=450$ K thermal fluctuations reduce the coercive field from its static value $\mu_0 H_{\mathrm 0} = 3.6$~T to $\mu_0 H_{\mathrm c} = 2.78$~T. 
This gives a relative change of the coercivity by thermal fluctuations of 
$\Delta H_{\mathrm {fl}}(450 {\mathrm K}) = 0.23$.

\begin{table}[t]
\caption{Effect of a \Dy shell on the coercive field of a \Nd grain. The first column gives the thickness of a defect layer with $K=0$. The second column gives the thickness of the \Dy shell. $H_0$ is the coercive field taking into account the temperature dependence of the intrinsic materials parameters only. $H_{\mathrm c}$ is the temperature dependent coercive field including thermal fluctuations. The last column gives the relative change in the coercive field owing to thermal fluctuations.}
\label{table:dy}
\begin{tabular}{c c c c c c}
      \hline\hline 
      defect(nm) & shell (nm) & $T$(K) & $\mu_0 H_0$(T) & $\mu_0 H_{\mathrm c}$(T) & $\Delta H_{\mathrm {fl}}$ \\ 
      \hline
      0 & 0 & 300 & 5.89 & 4.97 & 0.16 \\
      0 & 0 & 450 & 3.58 & 2.62 & 0.27 \\
      2 & 0 & 300 & 3.84 & 3.23 & 0.16 \\
      2 & 0 & 450 & 2.44 & 1.80 & 0.26 \\
      2 & 4 & 300 & 5.81 & 4.97 & 0.14 \\
      2 & 4 & 450 & 3.60 & 2.78 & 0.23 \\  
      \hline
\end{tabular}
\end{table}

In order to understand the influence of the \Dy shell on the thermal stability of the coercive field, we calculate $H_0(T)$ and $H_{\mathrm c}(T)$ for different configurations: (i) a perfect \Nd grain, (ii) a \Nd grain with a  2 nm thick soft magnetic surface defect ($K=0$), and (iii) the above discussed \Nd core / \Dy shell grain. Table \ref{table:dy} summarized the results. 
At $T=450$~K the perfect \Nd grain shows a coercivity of $\mu_0H_{\mathrm c} = 2.62$~T. 
Adding a surface defect the coercivity reduces to $\mu_0H_{\mathrm c} = 1.8$~T. 
The \Dy recovers coercivity and compensates the loss in $H_{\mathrm c}$ caused by the defect: The coercivity of the core/shell grain with a surface defect is $\mu_0H_{\mathrm c} = 2.78$~T. 
This value is higher than the coercivity of the perfect \Nd without a defect. 
While there is no guarantee that such a diffusion shell fabricated experimentally will be continuous and of constant thickness, the key message is that if you can make a perfect diffusion shell you only need 4 nm to reach the target coercivity. We are able to make a prediction, using simulation, of how thick the diffusion layer needs to be to reach the required coercivity. 

It is important to validate our simulations by comparing to experimental results. 
Sepehri-Amin et al. \cite{Sepehri-Amin2013} found experimentally a coercivity of 1.5 T for ultrafine-grained anisotropic Nd–Fe–B magnets, where the grains are platelet-like in shape. 
The reduction in coercivity due to grain shape was recently investigated by Bance at al. \cite{bance_grain-size_2014}, where it was shown that, for a 50 nm grain diameter, an aligned cubic grain has a 0.85 reduction in coercivity with respect to a dodecahedral grain shape. 
Additionally, grain easy axis misalignment reduces coercivity in real magnets. Sepehri-Amin et al. suggest a grain misalignment of around 15 degrees in their sample. 
Using a Stoner-Wohlfarth model this gives an approximate further reduction in coercivity of 0.61 \cite{Stoner1948, bance2014influence}. 
Combining the two reduction factors from shape and misalignment to adjust our calculated value for $H_{\mathrm c}$ of the dodecahedral grain with a defect but no hard shell at 300 K we go from 3.23 T to 1.68 T, which agrees well with the experimental value.

\section{Conclusions}

We presented a micromagnetic scheme for the computation of the temperature dependence of coercivity in permanent magnets. In addition to the change of the coercive field through the change of the temperature dependent anisotropy field, thermal fluctuations cause a further reduction of $H_{\mathrm c}$. This relative change of the coercivity owing to thermal fluctuations is around 15 percent at room temperature and 25 percent at 450~K.

The temperature dependence of the coercive field was calculated for Dy diffused permanent magnets. The results show that a \Dy shell of 4 nm only is sufficient to compensate the loss in coercivity caused by soft magnetic defects.

\section*{Acknowledgement}
This paper is based on results obtained from the future pioneering program ``Development of magnetic material technology for high-efficiency motors'' 
commissioned by the New Energy and Industrial Technology Development Organization (NEDO). The authors would like to acknowledge funding support from the Replacement and Original Magnet Engineering Options (ROMEO) Seventh Framework Program (FP7).
\bibliography{BanceJOMshort}
\end{document}